\documentclass[useAMS,usenatbib,usegraphicx]{mn2e}
 
\title[Hierarchical Disk Galaxy Assembly]{Hierarchical Disk Galaxy Assembly as the Origin of Scatter in the $z \sim 1$ Stellar Mass Tully-Fisher Relation}
\author[Atkinson, Conselice, Fox]{Nicola Atkinson$^{1}$, Christopher J. Conselice$^{1}$\thanks{E-mail:
conselice@nottingham.ac.uk}, Nicole Fox$^{1}$ \\
$^{1}$University of Nottingham, School of Physics \& Astronomy, Nottingham, NG7 2RD UK }
\def\deg{$^{\circ}\,$}
\def\solm{M$_{\odot}\,$}

\def\kms{km s$^{-1}$}

\def\deg{$^{\circ}\,$}
\def\solm{M$_{\odot}\,$}

\def\kms{km s$^{-1}$}

\def\casgm20{CAS-G-M$_{20}\,$}
\def\m20{M$_{20}\,$}

\begin{document}

\date{Accepted ; Received ; in original form}

\pagerange{\pageref{firstpage}--\pageref{lastpage}} \pubyear{2002}

\maketitle

\label{firstpage}

\begin{abstract}

Recent observations of distant disk galaxies show that there is
little to no evolution in the relation between maximum rotation
speed and stellar mass at $z < 1.2$. There is however
a significant scatter between these two quantities whose origin
is uncertain.   We show in 
this paper that this scatter is at least partially the
result of galaxy merging, revealing that disk galaxy growth at $z < 1$ is
fundamentally hierarchical.  We carry this out by calculating
CAS (concentration, asymmetry, clumpiness) structural parameters using 
archival Hubble Space Telescope
imaging of 91 high-redshift disk galaxies at $0.4 < z < 1.0$ 
with robustly measured stellar masses and rotational maximum velocities 
taken from Conselice et al. (2005).  We separate our sample into two 
redshift bins divided at $z = 0.7$, and investigate deviations from the
stellar-mass Tully-Fisher relation in both the
M$_{*}$ and V$_{\rm max}$ directions, and how these correlate with
structural asymmetries.  We find a significant 
($> 3\, \sigma$) correlation between the residuals from the 
stellar-mass Tully-Fisher relation in both the M$_{*}$ and V$_{\rm max}$ 
directions, and high asymmetries.  This result
 holds after we remove
contributions from star formation and edge-on galaxies which can produce
higher asymmetries unrelated to merging.  While there are a few cases in 
which our disk galaxies have very large asymmetries, and are potentially 
involved in major mergers, in general these asymmetries are smaller than the 
major merger limit. It is therefore likely that these galaxies are forming 
hierarchically through minor galaxy mergers, which is also suggested by the 
constant slope and zero point of the stellar mass Tully-Fisher relation 
during the same epoch.

\end{abstract}


\section{Introduction}

In the standard cosmological model the mass density of the universe is 
dominated by collisionless dark matter.  Structure within the universe 
forms as a result of small density fluctuations, which induce gravitational 
clustering. In this picture, smaller objects collapse and undergo mergers to 
form larger objects. Dark matter halos, in which galaxies are believed to be
embedded, form in this way. These halos form gravitational potential wells, 
which collect gas necessary for star formation, creating a 
visible galaxy (e.g., Lacey \& Cole 1994). 

Much observational work has been carried out to aid our understanding of the 
history of mass assembly in galaxies to test these ideas. Whilst 
dark matter halos build up through hierarchical methods, governed 
primarily by density fluctuations in the early universe, the assembly of 
stellar content is far more complex.  It is clear that 
gaseous dissipation, the mechanics of star formation, and the feedback of 
stellar energetic output of the baryonic material of galaxies must all be 
considered in understanding stellar mass build-up, along with the assembly
of galaxies through merging and accretion of existing stellar systems.

There are two popular ideas for explaining how the bulk of stars
in galaxies are put into place, namely the monolithic collapse,  and 
the hierarchical model.  Both of these models predict how disk 
galaxies and ellipticals form. In the monolithic collapse scenario stars 
are concentrated into a bulge due to a rapid collapse of either gas, or 
existing stars, contained by its dark matter halo. 
This bulge can then acquire a disk by cooling of gas from the intergalactic 
medium (IGM) onto the spheroid, with star formation occurring as the gas 
clouds collide (Steinmetz \& Navarro 2002; Abadi et al. 2004). In this model, 
different galaxy 
morphologies form due to discrete differences in environment, and the 
amount of matter present.  In the hierarchical model, spiral bulges are a 
result of the merging of existing stellar and gaseous systems. These mergers
lose angular momentum through the ejection of material  leading to 
high stellar concentrations in the central regions, comprising of both 
existing stars and those formed out of the colliding gas.

Whilst there is evidence for massive galaxies forming hierarchically
from mergers, creating  elliptical galaxies and bulges, the formation process 
for disk galaxies remains largely unknown.     One way to test this is through
the Tully-Fisher relation between a galaxy's luminosity and its
maximum rotational velocity (V$_{\rm max}$), as well as the evolution of
this relationship, and  its internal scatter.
A more fundamental stellar mass Tully-Fisher (SMTF) relation also exists, 
whereby a galaxy's luminosity is substituted by its stellar mass 
(M$_{*}$).  Bell \& De Jong (2001) and 
Verheijen (2001) both found a tight correlation between maximum rotational 
velocity and stellar mass, and K-band luminosity, for disk galaxies in the 
local universe. 

To search for evolution in the SMTF relation Conselice et al. (2005a) examined 
the SMTF relation at $0.2 < z < 1.2$  with respect to the $z \sim 0$ 
relation (Bell \& de Jong 2001). Conselice et al. (2005a) found that the 
relationship between stellar mass and V$_{\rm max}$ does not evolve 
significantly up to $z \sim 1.2$.  Since stellar mass in disks at $z < 1$
must be increasing due to observed star formation, the conclusion from this 
work is that disk galaxy formation at $z < 1.2$ is hierarchical - gas is 
accreted along with dark matter in a manner which preserves the stellar mass 
TF relation.  There is also no evolution in the stellar mass-total halo 
mass relation at the same redshifts (Conselice et al. 2005a),  suggesting 
that the stellar and dark mass components of disk galaxies grow simultaneously 
throughout this period.  This result was later also found to be the
case by e.g., Flores et al. (2006) and Bohm \& Ziegler (2006).

Another feature of the stellar mass-TF relation at high redshift is that 
there is a significant
scatter from the best fit relation. The origin of this scatter
is a major question that has been addressed in several previous works 
(e.g., Kannappan, Fabricant \& 
Franx 2002; Kassin et al. 2007). The origin of the scatter in the TF
relation can be the result of different mass to light ratios, different
formation histories in terms of star formation and mass assembly, or
intrinsic to disks themselves. By examining the stellar mass Tully-Fisher
we are removing the mass to light ratio differences and thus can test
whether the scatter is due to formation histories and/or is intrinsic, or
otherwise due to observational errors.  For high-redshift galaxies Kassin et 
al. (2007) examined the internal kinematics of $z < 1.2$ galaxies 
with a range of galaxy morphologies, and found that the TF relation
is highly 
morphologically dependent. Non-disturbed spirals form a clear ridge-line in 
the TFR that follows the local relation. 
Almost all major mergers, disturbed, and compact galaxies have lower 
rotational velocities compared to the $z \sim 0$ values, producing a large 
scatter. A larger scatter in the TF relation is also found for 
galaxies in close pairs with kinematic disorders 
(Barton et al 2001), and for irregular galaxies (Kannapan et al. 2002). 

Recently,  Flores et al. (2006) related the deviations from the TF relation 
with kinematic properties of galaxies using Integral Field Unit (IFU)
observations.  All rotating pure disks, with regular rotation curves, were 
found to lie along the $z \sim 0$ line. The majority of galaxies with
perturbed kinematics were found within the 3 $\sigma$ scatter quoted by 
Conselice et al (2005a), and galaxies with very complex kinematics 
deviated the most.  These galaxies 
deviate from the stellar mass TF relation due to too high or low 
rotational velocities, at a given stellar mass, produced by disturbed
 kinematics. If these eventually 
form well ordered disk  galaxies and move onto the $z \sim 0$ line, this 
evolution may result from their stars and gas settling into more 
circular orbits, hence increasing, or decreasing the measured V$_{\rm max}$. 

Conselice et al. (2005a) also found that many disk galaxies deviate 
strongly from the relation between
stellar mass and maximum rotational velocity. One hypothesis proposed
in Conselice et al. (2005a) and tested in this paper is that these galaxies 
are undergoing accretion of intergalactic matter, or satellite galaxies, 
which increases both the baryonic and stellar mass at roughly
the same rate. We test this hypothesis in this paper through an analysis 
of the CAS parameters for a sample of 91 disk galaxies used in the
Conselice et al. (2005a) study. If disks are forming hierarchically
from stellar+gaseous mergers then the systems which deviate
the most from the stellar mass TF relation should display
signatures of recent merging or accretion activity in their CAS
parameters.  

This paper is organised as follows. In \S 2 we present our sample of galaxies,
the data we use to perform this analysis, and the techniques
used to derive evolution. \S 3 presents the results of our analysis, \S 4
is a discussion of our results, and \S 5 is a summary of our findings.  
Throughout this paper we use the standard cosmology
of H$_{0} = 70$ km s$^{-1}$ Mpc$^{-1}$, and 
$\Omega_{\rm m} = 1 - \Omega_{\lambda}$ = 0.3, and a Chabrier
IMF for our stellar masses measurements, unless otherwise noted.

\section{Sample and Methods}

\subsection{Data and Sample}

The sample used in this analysis is taken from Conselice et al. (2005a), and
initially consisted of 101 disk galaxies imaged with the Hubble Space 
Telescope (HST) with rotation curves measured from the Keck Telescope, and 
for which we have deep near-infrared imaging. These disks were selected to 
have 
inclinations $> 30$\deg in order to remove nearly face-on systems so that 
rotation curves could be measured. However, no upper limit was set to the 
inclination, thus it is possible that dust lanes in edge-on galaxies are
present.  Our sample covers the redshift range $0.2 < z < 1.2$, enabling 
any evolution
in structural parameters to be seen over half the age of the universe. For
the purpose of image analysis, 
all galaxies were chosen to have no obvious interacting companions, or 
obscuring foreground stars. 

\begin{figure*}
 \vbox to 160mm{
\includegraphics[angle=0, width=140mm]{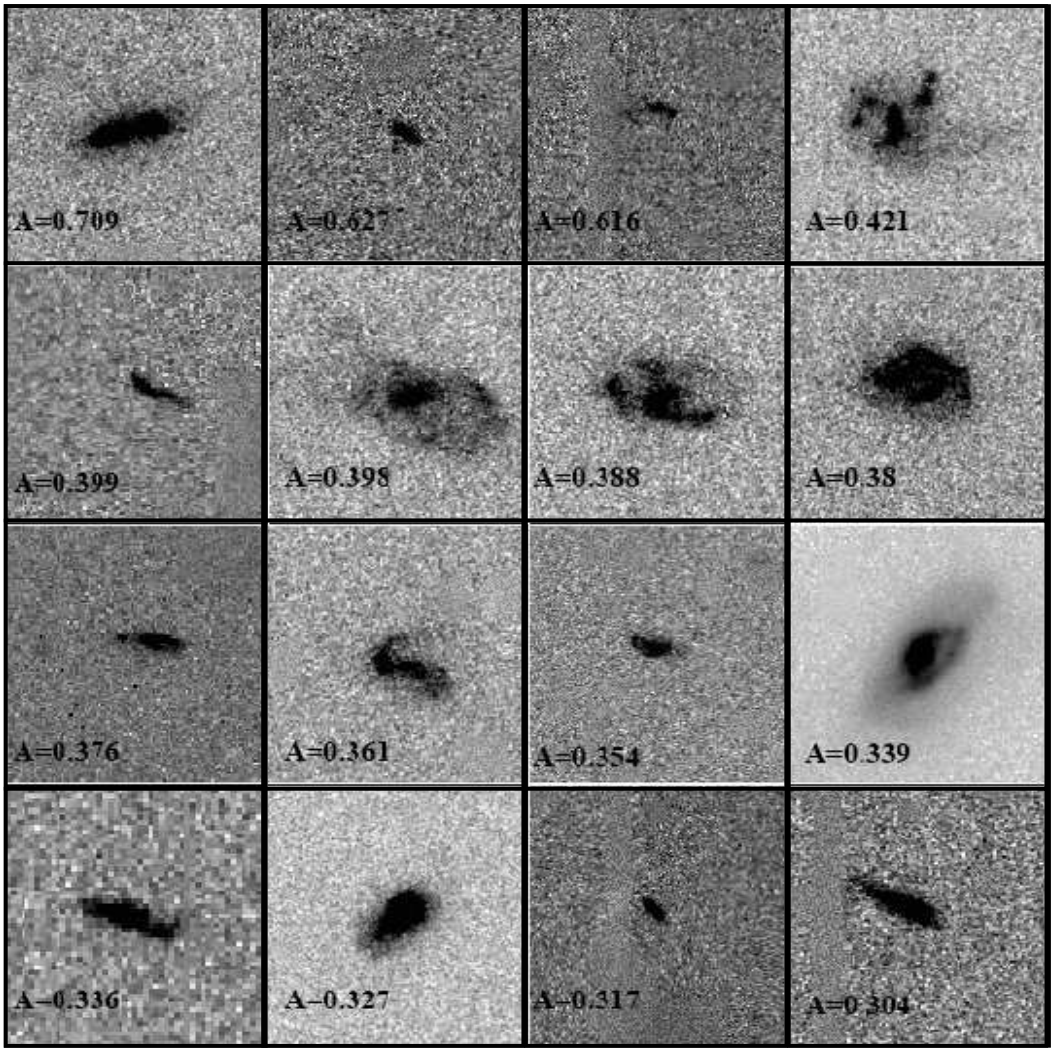}
 \caption{Galaxies with high asymmetries, $A>0.3$.  The asymmetry ($A$) is
indicated in the bottom left of each image. This Hubble Space Telescope
imaging is taken with the ACS and WFPC2 camera in the F814W band.}

} \label{sample-figure}
\end{figure*}

High resolution images of  these galaxies 
were taken from two sources, archival HST Advanced 
Camera for Surveys (ACS), and HST Wide Field Planetary Camera 2 (WFPC2)
imaging.  After removing systems that we could not use,  the total galaxy 
sample used in this project consisted of 91 disk galaxies. 
The redshift range was also reduced to $0<z<1$ to avoid strong morphological
k-corrections that can result in ambiguous results 
(e.g., Taylor-Mager et al. 2007).

The ACS Wide-Field Channel (WFC) allows deep, wide-field survey capabilities 
from the visible to NIR wavelengths.  
This difference however, should not pose a  problem when determining CAS 
parameters for these galaxies, since the $C$ and $A $
indices are sensitive to bulk structures, and are therefore 
remarkably stable with respect to resolution (Conselice 2003). The 
clumpiness parameter is more sensitive to resolution, and thus has a 
larger uncertainty in the WFPC2 images. 
The ACS WFC has a FOV of 202\arcsec\, and an optimal resolution of 
approximately 0.049"/pixel.  The WFPC2 images 
comprise data from four cameras, the Wide-Field cameras (WF2, WF3, WF4),
and the planetary camera (PC). The WF2, WF3 and WF4 have optics 
that are essentially the same, each having an  FOV corresponding to a 
resolution of 0.0996"/pixel.  

A total of 52 of the galaxies within our sample are found in HST 
images taken by the ACS Wide-Field Camera (WFC) in the Groth
Strip area of the DEEP survey (Vogt et al. 2008, in prep). The remainder of the
images are taken from archival WFPC2 programs including imaging of
the Hubble Deep Field-North and the Canadian-France Redshift Survey (CFRS). 
These images were available with their corresponding segmentation and weight 
maps, needed for determining the CAS parameters of a galaxy (e.g., 
Conselice et al. 2007c, in prep).  The images we use are taken within the
F814W (I-band) filter and the F606W (V-band) filter.

\subsection{Galaxy Rotation Curves}

The estimated disk galaxy V$_{\rm max}$ values used in the project were 
obtained by Vogt et al (1996, 1997, 2008) and Conselice et al. (2005a) 
using moderate-resolution 
spectroscopy from the Keck 10-m ground-based telescope with the
LRIS instrument (Oke et al. 1995). Spatially resolved 
spectra of these disk galaxies at $0.2<z<1.2$  were fit 
with a single (or double) Gaussian profile to each emission line (or doublet), 
most commonly the [OII] doublet, and the [OIII] and H$\beta$ lines 
(Vogt et al. 1993).   The position, amplitude, and width of the lines were 
found from best-fits to single or double Gaussian profiles, and the S/N 
ratio was calculated within these limits. The central 
wavelengths of these profiles were used to construct velocity rotation 
curves, where the Gaussian fit met minimum requirements in height and 
width of 5 $\sigma$ and 3 $\sigma$, respectively. However, the typical 
value was 10 $\sigma$ for both amplitude and width (Vogt et al. 1996). 

\begin{figure}
 \vbox to 120mm{
\hspace{-0.5cm} \includegraphics[angle=0, width=90mm]{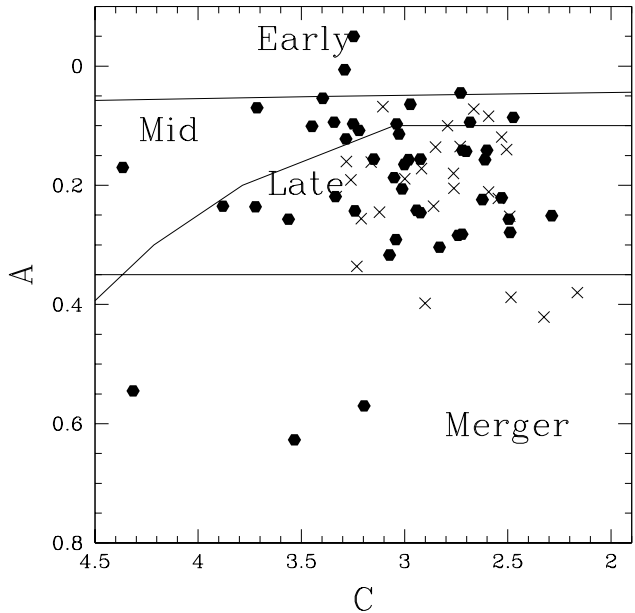}
 \caption{The concentration-asymmetry diagram for our sample of
disk galaxies in the observed F814W-band. Solid symbols are for
galaxies at $z < 0.7$ and crosses are for those at $z > 0.7$.  As can be 
seen, the vast majority of our sample of disks are found within the region of
CAS space occupied by late-type or early-type disk galaxies. There
are however clearly some examples of disk galaxies which appear to
have very high asymmetries.}
} \label{sample-figure}
\end{figure}

Determining the rotation curves for high redshift galaxies cannot be as
accurately done as for local galaxies. The sizes of the disk galaxies within 
our sample are typically just slightly larger than the size of the seeing, and 
the width of the spectroscopic slit. Therefore, wavelength shifts in the 
observed emission lines from which we measure V$_{\rm max}$ represents a 
combination of the spatial distribution in the disk's velocity and the 
emission line surface brightness. To correct for this Vogt et al. and
Conselice et al. (2005a) adopted a simple exponential model for each 
galaxy, matching inclination and orientation relative to the slit from
measurements from HST images. The rotational 
velocities from such models are assumed to have a linear form, rising 
out to a maximum at 1.5 times the disk scale length, R$_{\rm d}$, and then 
remaining flat at V$_{\rm max}$. The seeing was taken into account by 
convolving the model with an appropriate Gaussian, and the model emission 
lines were fit identically to the spectral data. Iterative adjustments were 
made to the rotational velocity of each galaxy model until the simulated 
and observed emission lines matched. 

Factors such as a misalignment of the slit with the galaxy major axis, and 
varying inclination are also considered when performing this simulation, as 
these also contribute to errors in the estimation of V$_{\rm max}$. It was 
also necessary to consider the relatively large uncertainty in the scale
length (R$_{\rm d}$) when estimating V$_{\rm max}$. Conselice et al (2005a) 
measured the disk scale-length, R$_{\rm d}$, by fitting a two-component 
model to the surface brightness distribution of each galaxy. However, they 
found that the errors in R$_{\rm d}$ resulting from this method may be 
underestimated. An additional uncertainty from the fitted R$_{\rm D}$ values
must therefore also be considered.  In total, the derived measurement of
V$_{\rm max}$ is inferred with an associated uncertainty of 20\%-30\%
(Conselice et al. 2005a). 

Whilst the above method works adequately for a well formed disk with circular 
orbits throughout, if a galaxy is undergoing a major or minor merger, 
the V$_{\rm max}$ 
estimate could be either higher or lower than expected for 
non-merging disks. In the beginning stages of a merger, the rotational 
motion of the two component galaxies may remain intact. However, at higher 
redshift, the components may not be spatially resolved. Only the spectra from 
the outer most parts of this combined system are considered when fitting the 
simple exponential profile, hence increasing the estimation of V$_{\rm max}$. 
Conversely, in the central stages of a merger, rotational velocity could be 
largely destroyed, hence lowering the  V$_{\rm max}$ estimate.   
Consequently, both positive and negative deviations in V$_{\rm max}$ 
from the $z \sim 0$ TFR could indicate mergers (see \S 3.2.1).

\subsection{Stellar Masses}

The stellar masses used within our sample are the same as those
calculated by Conselice et al. (2005a).  These stellar masses were determined 
using 
standard methods explained in e.g., Brinchmann \& Ellis (2000), Bundy 
et al. (2005, 2006) and Conselice et al. (2007a).  This method for calculating 
stellar mass combines near-infrared (NIR) luminosities and optical 
photometry of galaxies with well-known redshifts to estimate 
stellar masses based on fitting the observed SED to different star formation 
models.  We utilise models from Bruzual \& Charlot (2003) with a Chabrier
IMF to compute our stellar masses (see also Conselice et al. 2007). Unlike 
other methods of determining stellar mass of galaxies, this has no bias 
against galaxy type or orientation. The bands we used in Conselice 
et al. (2005a) to calculate the stellar masses are the K-band 
(2.2 $\mu$m), and the optical bands as observed with the Hubble Space 
Telescope.

\subsection{CAS Parameters}

We utilise the CAS (concentration, asymmetry, clumpiness) parameters 
(Conselice et al. 2000a,b; Bershady et al. 2000; Conselice 2003) to 
determine the morphological and structural state of our disk galaxy
sample.  CAS parameters provide classifications in terms of the 
concentration of light ($C$), asymmetry ($A$), and the clumpiness ($S$). 
Each of these parameters, and the apparent total magnitudes, are defined 
within the Petrosian radius aperture of,

$$R_{\rm Petr} = 1.5 \times r(\eta = 0.2),$$

\noindent where the dimensionless parameter $\eta$, is an intensity ratio of 
the enclosed light as a function of radius (e.g., Bershady et al. 2000).  
More than 99\% of the light is contained within this radius for most
galaxy profiles, and the light within the radius is the closest estimate to 
the true total magnitude, assuming a perfect Gaussian light profile 
(Bershady et al. 2000).  This radius is dependent only on the surface 
brightness within a given radius, and no assumptions are made about the shape 
of the galaxy's light profile. Below we briefly describe the CAS parameters
and how they are calculated within this radius.

\subsubsection{Concentration}

The concentration index (C) is a measure of a galaxy's integrated light 
distribution and is defined by Bershady et al. (2000) and Conselice (2003) 
as,

\begin{equation}
C = 5 \times {\rm log} \left(\frac{r_{80}}{r_{20}}\right),
\end{equation}

\noindent where $r_{80}$ and $r_{20}$ are the radii inside which 
80\% and 20\% of the light is contained, respectively. The
concentration of light correlates with the bulge to total light flux ratio 
of galaxies, and their masses (Conselice 2003). Thus concentration relates to a
galaxy's formation history in a broad sense. For disk 
galaxies, concentration is found between $2.5 < C < 4$, and 
galaxies with low central surface brightness generally have the lowest light
concentrations. 
This also corresponds to those galaxies with low internal velocity 
distributions on average (e.g., Bershady et al. 2000; Conselice 2003).

\subsubsection{Asymmetry}

Conselice et al. (2000a,b) and Conselice (2003) define the asymmetry index 
($A$) by rotating a galaxy image by 180\deg about its centre, and 
subtracting the flux values at each pixel from those of its pre-rotated 
image, $I_{0}$. The intensities of the absolute value of the residuals of this 
subtraction are then summed over all pixels, and are compared with the total 
galaxy flux. The basic mathematical definition of the asymmetry is,

\begin{equation}
A = {\rm min} \left(\frac{\Sigma|I_{0}-I_{180}|}{\Sigma|I_{0}|}\right) - {\rm min} \left(\frac{\Sigma|B_{0}-B_{180}|}{\Sigma|I_{0}|}\right),
\end{equation}
	
\noindent where the $B$-component is a correction for the background noise. 
Slightly negative $A$ values may be caused by the background correction if 
the galaxy is non-disturbed and compact with small residual values. 
Asymmetry is sensitive to any feature which produces distorted light 
distributions. These include galaxy interactions/mergers, and dust lanes 
occurring in edge-on disk galaxies (Conselice et al. 2000a). Star formation 
causes small scale 
disturbances, seen as high frequency information in an image, which
will also produce asymmetries.

Star formation however does not cause as high of an 
asymmetry signal comparable to the other dynamically induced
features. The asymmetry can also be affected slightly by the effects of 
projecting a 3-D galaxy onto a 2-D image, especially at very high 
inclinations (Conselice et al. 2000). A high asymmetry, by itself, 
therefore cannot be used to uniquely identify a merger. 

N-body models of active ongoing galaxy mergers also show that 
systems that have recently undergone a merger, or those at the beginning or 
end, of a major merger, may not appear highly asymmetric. For this reason the 
asymmetry index is not sensitive to all phases of merging, and 
Conselice (2003, 2006) show that in a given sample the total number of 
galaxies undergoing major mergers would be underestimated by roughly 
a factor of 
$\sim 2$ if only using the asymmetry parameter. However, no non-mergers have 
extremely high asymmetries unless they are at high inclinations, 
where dust lanes and projection effects become important.

\begin{figure}
 \vbox to 125mm{
\hspace{-0.5cm} \includegraphics[angle=0, width=90mm]{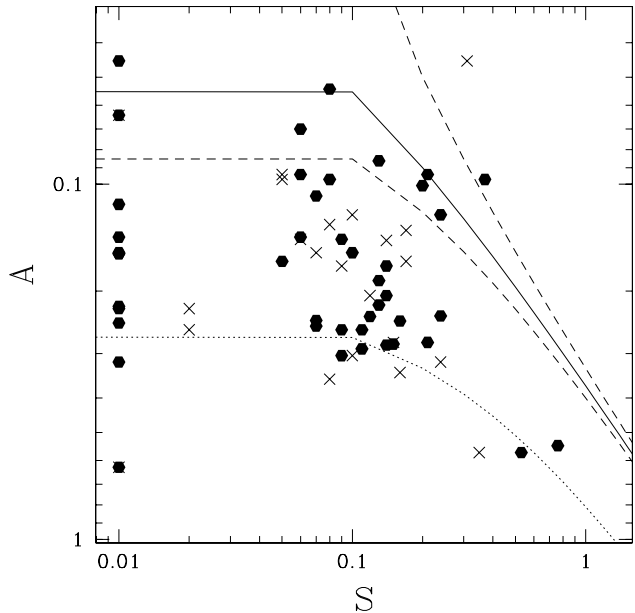}
 \caption{The clumpiness-asymmetry diagram for our sample of
disk galaxies in the observed F814W-band. The solid points are
for $z < 0.7$ galaxies and the crosses are for those at $z > 0.7$.  
The solid line on
this diagram shows the location of nearby normal galaxies as
found at $z \sim 0$, with the 3 $\sigma$ scatter from this relationship
shown as the dashed lines.  The dotted line shows the derived relation
for systems which deviate strongly from the relation between $A$ and
$S$.  The large number of points at S = 0.01
are for those galaxies which are too small to have a correctly
measured $S$ parameter (Conselice 2003).}
} \label{sample-figure}
\end{figure}

\subsubsection{Clumpiness}

The clumpiness parameter ($S$) is a measure of the patchiness of a galaxy's 
light distribution. It is defined by Conselice (2003) as the ratio of the 
amount of light contained in high frequency structure (within the defined 
galaxy aperture) to the total amount of light in the galaxy. In order to 
compute this quantity, the original galaxy image is smoothed by a filter of width
$\sigma$ to reduce its effective resolution, thus removing all high 
frequency  information. This smoothed image is then subtracted from 
the original image  to obtain a residual image leaving only the 
high frequency components. This flux is then summed over all pixels and a 
background brightness value is subtracted. Mathematically, clumpiness ($S$) 
values are obtained through the formula,
	
\begin{equation}
S = 10 \times \left[\left(\frac{\Sigma (I_{x,y}-I^{\sigma}_{x,y})}{\Sigma I_{x,y} }\right) - \left(\frac{\Sigma (B_{x,y}-B^{\sigma}_{x,y})}{\Sigma I_{x,y}}\right) \right],
\end{equation}

\noindent with any negative differences forced to zero. The inner 
parts of each galaxy are also not considered in the computation of $S$
since they often contain high frequency information unrelated to 
stellar light distributions (Conselice 2003).   Since 
disturbances due to star formation appear as high frequency information, a 
large amount of star formation results in high $S$ values. High clumpiness is 
therefore more prominent in star forming galaxies, while $S \sim 0$ for 
ellipticals which contain little star formation and appear smooth. 

\subsubsection{CAS Computations and Meaning}

Before performing our structural parameter measurements foreground stars, 
background galaxies, cosmic rays, hot pixels, and light gradients are
removed from the images.  Some of these features are removed by the SExtractor
segmentation maps,  while those missed by this process are cleaned by 
hand.  Many of the galaxies in our sample are also found in more that one 
HST archival image. We use the image with the highest signal to noise, and 
the lowest amount of contamination from other light sources, and those in 
which the galaxy is not close to the edge of the CCD chip.

As we use both I-band (F814W) and V-band (F555W) images in our analysis, we 
can compare directly how the CAS values change between these two wavelengths.
We find that the C, A and S parameters in the V-band were larger by
roughly 3\%, 9\% and 109\%, respectively, compared to the I-band values. 
However, the $C$ parameters for the V-band are only higher than the 
I-band images for 66\% of the images. Similarly, for the $A$ and $S$
parameters the V-band images have parameters higher than the I-band images 
for 58\% and 76\% of the images, respectively.

We apply the CAS structural parameters to decipher 
any structural evolution from low to high$-z$ in our disk galaxy sample, and
to determine how this correlates with the scatter from the SMTF relation. 
It is therefore important to consider how measurements of these
parameters change as galaxies become less resolved and fainter due to 
cosmological effects.  To address this, Conselice (2003) simulated 
nearby galaxies  at various higher redshifts to determine the effects
of redshift on measuring galaxy structures.  Conselice (2003) find
that the asymmetry and clumpiness 
indices decline, on average, as a function of redshift, and the 
concentration increases slightly.   The amount of this change
using 1-orbit HST images is roughly $\delta A = -0.03$, $\delta S =$ -0.25,
and $\delta C$ = 0.1 at $z \sim 1$.

In addition to the CAS measurements we also visually classified all galaxies,
and found disks that appear to be disturbed or compact, and 
compared these to the calculated parameters. The most disturbed, highly
asymmetric galaxies are shown in Figure~1. We find that galaxies with
high $A$ values in general look by eye more disturbed.
All galaxies with low asymmetries appear as symmetric, and sometimes 
compact, disks.

More 
than 80\% of the galaxies with $A>0.3$ are found within the ACS images, with 
three WFPC2 images in this group. 
The larger percentage of ACS images with high asymmetry values suggests that 
the image resolution affects the calculation of this parameter. However, 
calculation of the asymmetry parameter is mostly dependent on 
large scale-structures, which are not easily lost through 
reduced resolution. Therefore it is 
possible that there could simply be few high asymmetry galaxies in the 
WFPC2 sample and upon inspection, there were none in which a galaxy appeared 
highly asymmetric in its large scale structure.

Conversely, all galaxies with $A<0.05$ were found in the WFPC2 images. 
We find that 45\% of the WFPC2 images have zero clumpiness values in both 
bands, compared to only 4\% in the ACS images.   The $S$ parameter 
is dependent on high frequency information, which is lost in low 
resolution imaging. For the lower resolution WFPC2 images, the galaxy 
subtends fewer pixels and, combined with the use of only the central region 
for computation, there is a very limited amount of information available 
from which to deduce meaningful clumpiness values.

In several cases a galaxy could not be measured within the CAS system, and
was excluded from the sample, since it was too contaminated by light from a 
neighbouring source. Often this contamination was the result of nearby 
sources that could not be completely removed successfully. Consequently, 
CAS parameters are calculated 
for a total of 91 disk galaxies.

\begin{figure*}
 \vbox to 100mm{
\includegraphics[angle=0, width=180mm]{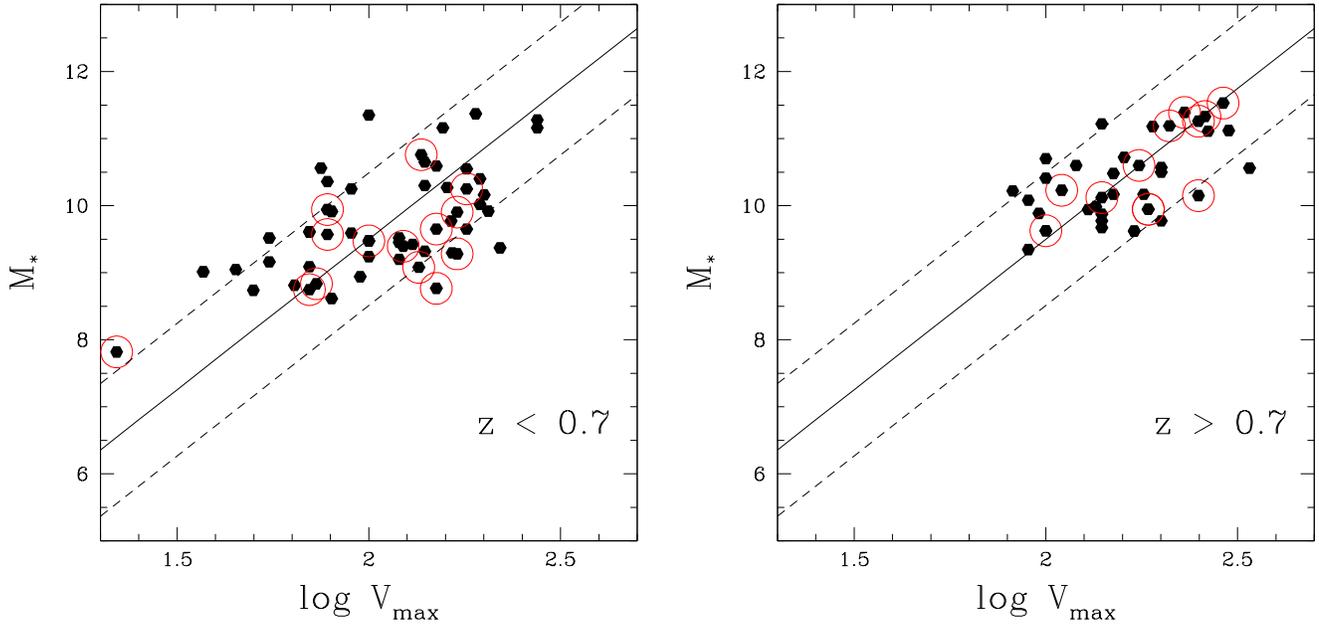}
 \caption{The Tully-Fisher relation for galaxies within our sample at $z<0.7$.
 The solid line shows the $z \sim  0$ relation given in Conselice et 
al. (2005a), and the dashed-lines show the 3 $\sigma$ scatter in this
relationship. The error-bars for these points are discussed in detail
in Conselice et al. (2005a).   M$_{*}$  is measured in 
solar mass units (\solm), and V$_{\rm max}$ is measured in \kms. The right
hand side shows the TF relation for galaxies within our sample at $z>0.7$.
Galaxies which are circled on this diagram are for those which have asymmetry
values $A > 0.25$.}
} \label{sample-figure}
\vspace{1cm}
\end{figure*}

\subsubsection{Identifying Morphological Peculiarities}

Asymmetry and clumpiness parameters reveal active evolution in the form of 
interactions/mergers and star formation, respectively.   We thus use the 
CAS parameters to determine how galaxy formation processes correlate with  
deviations from the SMTF relation.   Given the postulate that galaxies in 
our sample are deviating from the SMTF relation due to mergers, we would 
expect that galaxies which 
deviate highly will contain higher asymmetries.  Before
we can test this, we must examine the methods used to determine
the presence of galaxies undergoing active merging.

As described in Conselice et al. (2000a,b) and Conselice (2003), a
selection of $A > 0.35$ for nearby galaxies will give a nearly 
clean selection of active ongoing major mergers, uncontaminated by normal 
galaxies.  Another way to identify 
mergers, or galaxies strongly interacting, is to examine the relationship 
between the $A$ and $S$ parameters. Conselice (2003) 
found a relation for non-merging galaxies from all morphologies, given
by,

\begin{equation}
A=(0.35\pm0.03) \times S + (0.02\pm0.01).
\end{equation}	

\noindent Deviations from this fit, due to increased asymmetries, are  
caused by distortions in a galaxy's stellar light distributions, 
possibly indicating mergers.  This is certainly the case for
nearby galaxies in rest-frame optical light (Conselice 2003).
However, since systems that have recently finished a 
merger, or are at the beginning of a merger, will have settled in their 
distribution, these deviations will not locate all ongoing mergers
(Conselice 2006).  
Galaxies involved in minor mergers should also deviate slightly from 
the A-S relation compared with non-mergers, since their asymmetries 
should also be higher than average (Conselice 2006).

\section{Results}

\subsection{CAS Volume}

The CAS parameters for our sample of galaxies are compared to the CAS volume 
classification in Figures~2 \& 3. Since our sample was chosen to only  contain 
disk galaxies, we expected that our galaxies would lie within the bounds 
of the CAS disk morphologies: early-type disks, late-type disks, and 
edge-on disks. This 
implies that the majority of the calculated parameters would lie within the
region $2.7<C<4.3$, $0.03<A<0.28$, and $0<S<0.65$, with the 1 $\sigma$ 
variation in the classification system taken into account (Conselice 2003).

\begin{figure}
 \vbox to 105mm{
\includegraphics[angle=0, width=90mm]{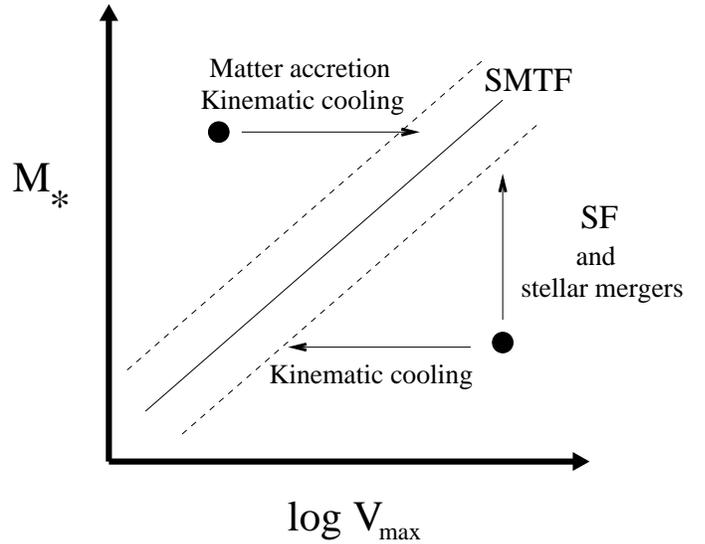}
 \caption{Graphical representation for how galaxies which deviate
from the SMTF relation can evolve onto it.
Shown by lines with arrows are the direction in this plot a galaxy
would evolve due to kinematic cooling after a merging event which can both
lower and increase the value of V$_{\rm max}$.  Also shown are how with
new star formation (SF), and stellar mass accretion events, the 
value of M$_{*}$ can increase over time. }
\vspace{-3cm}
} \label{sample-figure}
\end{figure}

We find that 70\% of the disks in our sample have parameters which 
placed them within the CAS region enclosing disks and spiral galaxies. 
The average redshift is $z=0.49$ for those galaxies classified as late-type 
disks within CAS, and $z=0.68$ for early-type disks, suggesting that these 
could be correctly identified, as more massive disks are selected at the 
higher redshifts (Conselice et al. 2005a).  We further find that 18\% of 
our sample have asymmetries $A>0.3$, and the values correlate by eye to the 
amount of disk distortion, as seen in Figure~1. 

We compare the asymmetry index ($A$) with the clumpiness index ($S$) 
for the $z < 0.7$ and $z > 0.7$ redshift bins in Figure~3.  
If we utilise the relationship for non-mergers we can infer 
if any galaxies outside the 3 $\sigma$ scatter could be mergers. 
However, due to resolution and signal to noise effects, the A-S relation 
is offset with respect to the $z \sim 0$ relation due to a lowered
resolution and signal to noise, and we have to 
consider this offset when using the A-S diagram.

\begin{figure}
 \vbox to 120mm{
\hspace{-0.5cm} \includegraphics[angle=0, width=90mm]{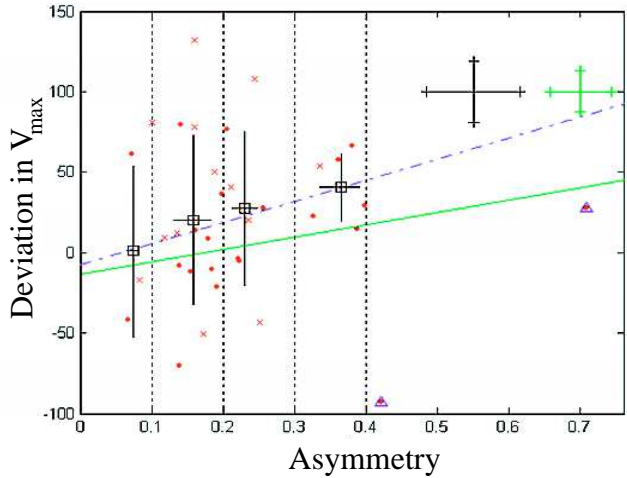}
 \caption{The $A$ parameter calculated from galaxies in both ACS and WFPC2
I-band images (F814W) plotted with the 
deviation in $V_{\rm max}$ from the $z \sim 0$ relation. The small
solid points are for galaxies from the ACS imaging, which the crosses
are for the WFPC2.  
Galaxies above the $z \sim 0$ line in Figure~4 have positive residuals and 
those below have negative. The solid green line shows a best fit with no 
exclusion of points. The blue dashed line shows the best fit, where the 
high asymmetry points indicated by blue triangles have been excluded. Black 
squares 
indicate the average values of all points contained within each A-bin 
marked by the dotted black lines. Black crosses mark the corresponding 
standard deviations in the scatter for each group of points.  Black and 
green error-bars indicate ACS and WFPC2 errors respectively. }
} \label{sample-figure}
\end{figure}

The A-S relation for non-mergers in Figure~3 is plotted as the solid line. 
After accounting for redshift corrections,  we find that $\sim$ 50\% of 
our galaxies lie outside of the 3 $\sigma$ 
scatter. We exclude any galaxies which lie within the 3 $\sigma$ scatter 
of the relationship for non-merging galaxies, and plot a best fit line 
for the remaining points, weighted by the parameter errors, to calculate
the relation between $S$ and $A$ for merging galaxies. For the the 
WFPC2 images, this was fit as,

\begin{equation}
A = (0.7 \pm 0.2) \times S + (0.23 \pm 0.03),
\end{equation}

\noindent where the errors quoted are the 95\% confidence bounds. Using 
the ACS images, the merging relation is very similar,

\begin{equation}
A = (0.6 \pm 0.3) \times S + (0.21 \pm 0.03),
\end{equation}
			
\noindent as shown in Figure 3.   Both of the above fits (Eq 5 \& Eq 6) 
have a RMSE of 0.09, showing good relations. The overlap in the error-bounds 
of the above relations indicate that the wavelength and instrument 
dependency in the CAS parameters must alter the A and S parameters in a 
systematic way. 

A larger number of merging galaxies are expected at 
higher redshift (e.g., Conselice et al. 2003). Since merging systems have 
higher $A$ 
values, we examine the correlation between redshift and the 
asymmetry index.  This was 
done by examining the relationship between asymmetry and redshift for only 
galaxies with calculated asymmetries $A > 0.25$. Based on this,
we find that there are a greater number of highly asymmetric galaxies 
at higher redshift. There are very few galaxies 
with $A>0.25$ at redshift $z<0.4$. We also investigate how the 
concentration parameter varies as a function of redshift, finding no 
significant correlation.

\begin{figure}
 \vbox to 100mm{
\hspace{-0.5cm} \includegraphics[angle=0, width=90mm]{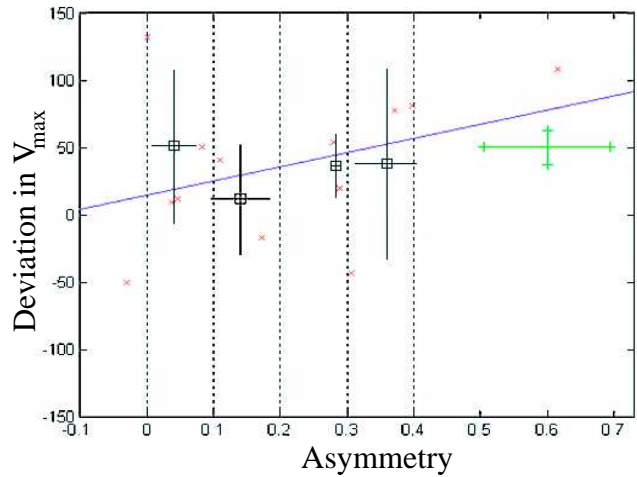}
 \caption{The asymmetry parameter calculated from V-band WFPC2 images plotted
against the deviation in V$_{\rm max}$ from the $z \sim 0$ relation. Galaxies above 
the $z \sim 0$ line in Figure~4 have positive 
residuals and those below have negative residuals. The solid blue line 
shows the best fit. The black squares indicate the average values as 
described in Figure~6.  Errors are also as described in the 
Figure~6 caption.}
} \label{sample-figure}
\end{figure}

\subsection{Structure-Tully-Fisher Relation at $z < 1.2$}

\subsubsection{Motivation}

As we have described in the introduction and in \S 2.2,
correlating the deviations from the SMTF with galaxy structure
may reveal the origin of the large scatter in SMTF relation at
$z < 1$.  Before we analyse how the scatter in the SMTF relates
to galaxy structure, it is important to understand the motivation
behind this test.  Figure~4 shows the SMTF relation as observed
at $z < 1$, which was previously published in Conselice et al. (2005a).
As can be seen, there is significant scatter in the relation
between V$_{\rm max}$ and M$_{*}$, although there is a clear
general relationship between these two quantities.

Figure~5 gives a graphical model for how galaxies which do not fall
along the SMTF can evolve onto it over time.  The processes by which
a galaxy can move onto the SMTF relation include kinematic cooling from
a past galaxy merger, as well as an increase of stellar mass
from star formation, and through the accretion
of existing satellite galaxies.  Whether these processes can occur 
depends on if a galaxy is located above the SMTF relation or
below it.  If a galaxy is located above the relation, then the only
way it can reach the relation is through the growth of dark matter without
accompanied star formation, and/or the kinematic cooling of a galaxy
after it undergoes a dynamical event with another galaxy, either through
a merger or an interaction of some type.

For those galaxies below the relation, the evolution onto the SMTF
can occur either through stellar mass evolution through star formation,
or accretion of satellite galaxies, and/or through kinematic cooling
after a merging or interaction event. In all but one case, the evolution
onto the SMTF relation involves hierarchical growth.  We can rule out
interactions with other galaxies that do not involve merging, as our
disk galaxy sample do not have any nearby companions. Therefore, any evolution
on the SMTF relation in the V$_{\max}$ direction involves either
kinematic cooling from a merger, or the accretion of matter from
the intragalactic medium.  Evolution onto the SMTF
from galaxies above the relation therefore must involve some type of dynamical
accretion or merging event.

On the other hand, it is possible that galaxies below the SMTF reach
the relation over time through star formation.  The other possibility
is that the scatter in the SMTF relation at these redshifts is purely 
due to observational
errors.  By correlating this scatter with galaxy structure, and
determining the significance of any trends, we can make this 
distinction.

\subsubsection{The Stellar Mass Tully-Fisher}

To examine how structure correlates with the scatter in the stellar
mass TF relation, we split our galaxy sample into two redshift bins, 
$z<0.7$ and 
$z>0.7$.   We determine deviations from the SMTF based
on the best fit to the observed SMTF relation using our own data, and to the
$z \sim 0$ relation.  The zero point of 
the Bell \& de Jong (2001) $z \sim 0$ SMTF relation was fit such that
the standard deviation from the line was at a minimum. For the $z < 0.7$ 
systems this was found to be:

\begin{equation}
{\rm log}\, M_{*} = (4.49 \times {\rm log} V_{\rm max}) + 0.37,
\end{equation}

\noindent and for $z > 0.7$ the relation we fit is,

\begin{equation}
{\rm log}\, M_{*} = (4.49 \times {\rm log} V_{\rm max}) + 0.29.
\end{equation}

\noindent These derived relations are both within the error bounds of the 
local relation found by both Conselice et al (2005a), and Bell \& de 
Jong (2001).   This is consistent with no large scale evolution of the 
SMTF relation over the redshift range $0 < z < 1$. Hence, we will assume 
that our best 
fits in Eq 7 \&  8 describe the SMTF relations at their respective
redshifts.  If we utilise either the $z \sim 0$ relation, or these internal
fits, we find very similar results in what follows.

Galaxies which deviate by more than 3 $\sigma$ from the SMTF relation 
were visually 
inspected for disturbances, and the CAS parameters analysed for any 
trends. Over redshifts $0<z<1$, we found that deviating galaxies 
covered a large range of asymmetry values, $0<A<0.62$.  We also checked if 
any deviating galaxies appear as either face-on or edge-on disk
galaxy systems. Face-on galaxies could have inaccurate V$_{\rm max}$ 
estimates, and dust lanes may affect the asymmetries of edge-on galaxies. 
When looking at images of deviating galaxies, only one galaxy 
appears face-on, displaying a central bar structure. 

Deviations in  terms of V$_{\rm max}$ and M$_{*}$ from the best fit $z \sim 0$ 
relations (Eq 7 \& Eq 8) were calculated in each of our two redshift 
bins. Deviations are such that if the value is higher than predicted
by the SMTF relation, then the residual is negative.  
These deviations were compared with the asymmetry and clumpiness
parameters. First, we found no clear and obvious relation between 
highly deviating galaxies and the asymmetry parameter for galaxies 
at $z < 0.7$. However, Figure~6 and Figure~7 show the  
relation between V$_{\rm max}$ deviations from the SMTF relation, and the 
asymmetries of these galaxies
at $z > 0.7$, where we do see a significant correlation within
the I-band and V-band images, respectively. 

We fit a line to the V$_{\rm max}$ residuals, as a function of
asymmetry, weighted by the errors in the deviation in V$_{\rm max}$. 
This fit is shown in Figure~6 as a solid green line.  However, a better 
fit was obtained if the two high A galaxies, indicated with blue triangles, 
were excluded. The standard deviations in the scatter place the points 
on this new relation, which is shown as a blue dashed line. 
For galaxies in the V-band imaging, a weighted best-fit line was 
derived, shown as the solid green line in Figure~7. Although no points have 
been excluded, this relation is less reliable since there are only 13 
V-band galaxy images at $z > 0.7$. However, a positive 
correlation is seen in both Figure~6 and Figure~7.

We utilise Monte Carlo simulations to test 
the validity of these correlations, and to compare whether this increase
in asymmetry with SMTF residuals is significant given the large 
uncertainties in measuring M$_{*}$ and V$_{\rm max}$.  To carry out this
Monte Carlo simulation we place each data point randomly along
its V$_{\rm max}$ error-bar. This simulation was repeated 100,000 
times, and we find that the significance of the correlations are 
3 $\sigma$ and 1 $\sigma$, for the I-band and V-band deviations from the 
$z \sim 0$ line, respectively. The actual significance of these correlations
is likely even higher than this, as these errors likely have a normal 
distribution.

As explained in \S 2.2, both positive and 
negative deviations in V$_{\rm max}$ could indicate mergers. 
Therefore, we also studied the equivalent of Figure~6 and Figure~7 for the 
absolute residuals in V$_{\rm max}$ from the $z \sim 0$ line. No 
significant trend (1 $\sigma$) was seen in this data, showing that not all 
highly deviating galaxies have high A values. It also shows that on average
galaxies with a higher asymmetry have a larger positive deviation in 
V$_{\rm max}$ from the best fit SMTF relation.

\subsubsection{Contribution from Star Formation}

Since a high asymmetry index can be produced through both mergers/interactions
and star formation, we investigate whether the
correlation between the asymmetry and deviations from V$_{\rm max}$
are produced by star formation. To investigate this, a new parameter
$A' = A - S$ is computed for all galaxies in 
our sample.   This new parameter indicates in a sense
`true' asymmetries, with the contribution from star formation and 
dust removed by the clumpiness parameter which correlates with small-scale 
features in
a galaxy (Conselice 2003).  The relation between $A'$ and the SMTF relation
residuals
is shown in Figure~8.   We find no major change to the trend seen in 
Figure~6 between the asymmetry itself and the SMTF relation residuals, 
suggesting that 
this correlation is produced through large-scale features in a galaxy. 
The two galaxies with high asymmetries in Figure~6 also have high $S$ 
parameters, and their positions have moved significantly, and they
no longer deviate strongly from the trend.   The original asymmetry
values for these galaxies are thus clearly affected by star formation,
but the vast major of asymmetries are not.

Whilst deviations in V$_{\rm max}$ from the $z \sim 0$ TFR are useful for 
identifying galaxies which may have undergone kinematic changes, deviations 
in M$_{*}$ may indicate either star formation or stellar mass accretion 
(\S 3.2.1). Galaxies which have large positive deviations in M$_{*}$ 
resulting from a merger 
would be expected to have high asymmetry values. Such a relation can be 
seen in Figure~8. 

In general, galaxies that display a higher positive 
deviation in the SMTF relation residuals in M$_{*}$ space have higher 
computed asymmetries.   Monte Carlo 
simulations of this correlation shows that this result is significant at 
$>$ 3 $\sigma$. 
However, an exception to this is one galaxy which has a high 
positive deviation in M$_{*}$, despite its relatively low asymmetry 
($A = 0.16$). The asymmetry parameters measured from the V-band
images showed a similar trend to that seen in Figure~9, but due to the 
small sample size cannot be used to infer any significant relation.

Since mergers often induce star formation, we examine the relationship 
between clumpiness and the deviations in both V$_{\rm max}$ and M$_{*}$. 
However, no trend was seen at any redshift, partially due to the 
difficulty of measuring the clumpiness parameter.

\begin{figure}
 \vbox to 100mm{
\hspace{-0.5cm} \includegraphics[angle=0, width=90mm]{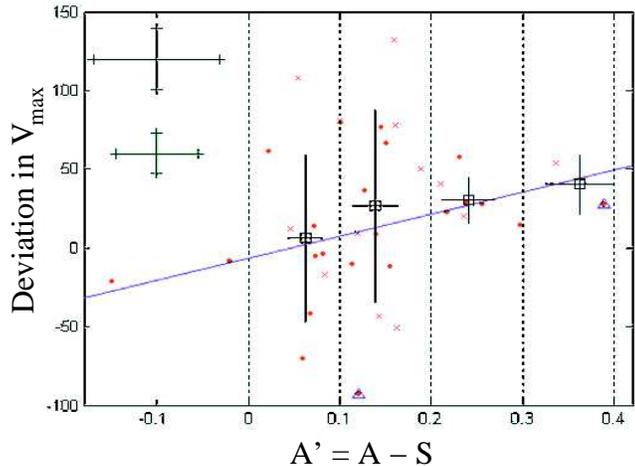}
 \caption{The $A'  = A - S$ parameter calculated from the I-band images 
compared with 
the deviation in V$_{\rm max}$ from the $z \sim 0$ relation.  
Blue triangles denote the two high asymmetry galaxies shown in Figure~6. 
The solid blue line 
shows a best fit. The error in $A'$ is from the combined errors in the
$A$ and $S$ parameters. All other features are as outlined in Figure 6.}
} \label{sample-figure}
\end{figure}

\section{Discussion}

\begin{figure}
 \vbox to 100mm{
\hspace{-0.5cm} \includegraphics[angle=0, width=90mm]{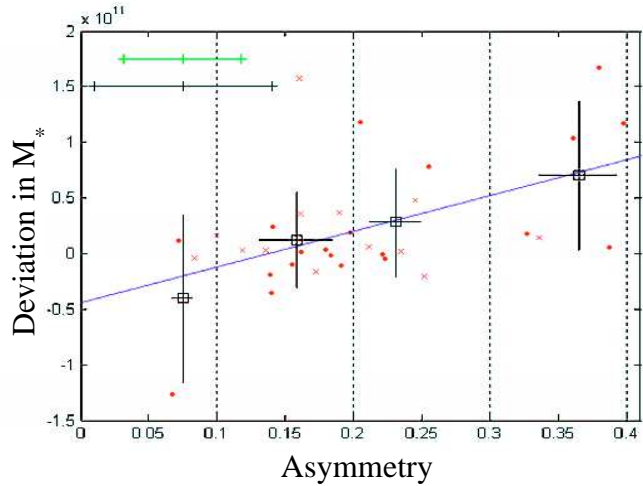}
 \caption{The asymmetry parameter calculated from the I-band images 
compared with 
the deviation in M$_{*}$ from the $z \sim 0$ relation. 
 Galaxies above the $z \sim 0$ line in Figure~4 have positive residuals and 
those below have negative residuals. The solid blue line shows a best fit. 
The black squares indicate the average values of all the points contained 
within each A-bin marked by the dotted black lines. The black crosses mark 
the corresponding standard deviations in the scatter for each group of 
points.  The green and black error-bars indicate the WFPC2 and ACS errors, 
respectively. }
} \label{sample-figure}
\end{figure}

We find in this paper that the deviations from the
SMTF relation in both the M$_{*}$ and V$_{\rm max}$ directions correlate
with galaxies with higher internal asymmetries.  We have also
found that disk galaxies at higher redshifts display on average higher
asymmetries. 

The number of mergers seen in our sample, based on the asymmetry
parameter, falls short by a factor of $\sim 2$ of the 22\% of 
mergers found occurring in disks at $z \sim 0.6$ by Flores et al (2006).  
This is consistent with the postulate in Conselice (2003) 
that in a sample of galaxies, the total number undergoing mergers would be 
underestimated by a factor of two when considering analysis of the asymmetry
parameter alone.  Since 
Flores et al (2006) used IFU measurements to look at perturbed 
galaxy kinematics their estimate of the number of mergers is thought 
to be more accurate than CAS parameters allow.  However, our sample is
three times larger than Flores et al. (2006) and covers a wider range
of redshifts.

The derived SMTF relation for our sample, at both $z > 0.7$ and $z < 0.7$, are 
similar to each other, and are consistent with no significant evolution of the 
SMTF relation over the redshift range $0<z<1$ (Conselice et al. 2005a).   
The origin of these deviations from the SMTF may arise from accreted gas, 
stars, and dark matter from the IGM in the form of minor mergers or smooth 
accretion.  These accretion mechanisms can increase the stellar mass with 
either pre-formed stars, or by inducing star formation. Since mass 
cannot be lost from a galaxy, evolution in this case must settle 
towards higher V$_{\rm max}$. This settling could be caused by 
circularisation of gas orbits or gaseous dissipation to the disk plane, 
ending with the growth of a gaseous disk. 

Deviations below the $z \sim 0$ SMTF relation are for disks with either
a lower M$_{*}$ and/or a higher V$_{\rm max}$ than expected. If 
these deviations are due to the early stages of a merger, the increased 
V$_{\rm max}$ may result from individual components
not being resolved in the ground based spectra. 
This causes incorrect V$_{\rm max}$ estimations, and a falsely decreased 
stellar to halo mass ratio. The deviation could also be 
caused by a disk galaxy which has not yet formed all its stellar components 
and the inferred stellar to halo mass ratio could be evolving.  Galaxies
found below the $z \sim 0$ line therefore would need to evolve in stellar
mass through accretion processes or star formation, so that this ratio 
would become that expected for the disk galaxy TF relation.  This reasoning 
suggests that the majority of scenarios for deviations both above and 
below the $z \sim 0$ line can be explained by the occurrence of mergers.

At $z>0.7$, we find that $\sim$ 70\% of disks with stellar masses 
M$_{*} > 10^{10}$ \solm  scatter to low V$_{\rm max}$, which shows a 
slight increase compared with those at $z<0.7$.   
For M$_{*} < 10^{10}$ \solm galaxies at $z>0.7$ there are no systems 
which deviate 
from the $z \sim 0$ line by more than 3 $\sigma$, likely an
effect of the sampling bias. Lower mass disks may have been missed at 
high redshift because they often are not bright enough to observe
rotation  curves. At lower redshift however, there are large deviations at 
both high and low V$_{\rm max}$. This suggests that lower mass 
galaxies at low redshift may be still forming.   

Interestingly, we do not find a significant trend between asymmetries
and residuals from the SMTF relation for galaxies at $z < 0.7$.
A  possible reason for this is that structural relaxation is 
quicker than the internal velocity relaxation after a merger. In such a case, 
the stellar structures are relaxed before V$_{\rm max}$, which would be the
case as the V$_{\max}$ values originate from larger radii where the
time-scale needed for dark matter in the halo to 
settle are longer than the stars at the centre of the disk.

Figure~6 shows that there is a positive deviation in V$_{\rm max}$ 
from the SMTF relation at higher $A$. This means that highly 
asymmetric galaxies have large scatter at lower V$_{\rm max}$ values. 
This supports the idea that these deviations are due to the 
central stages of mergers. In this case the galaxy appears asymmetric,
and has a 
random velocity distribution causing the lower V$_{\rm max}$ measurement. 

It is also apparent that deviations at higher V$_{\rm max}$ 
(negative residuals) most often have low asymmetry values. This could 
be explained if the galaxy is at the beginning stage of a merger where 
its asymmetry parameter has not 
yet been affected. An exception to this observation however, is one galaxy
which has a large deviation at higher V$_{\rm max}$, but a relatively 
high asymmetry of $A = 0.42$. We suggest that this galaxy 
fits within our hypothesis 
that deviations at higher V$_{\rm max}$ indicate the beginning stages of 
a merger.  However, in this case the merging components can be visually 
seen, producing a high asymmetry despite the 
structure of the individual components remaining intact. 

When examining deviations from the best-fit SMTF relation in M$_{*}$, we find
that, in general, larger deviations of higher M$_{*}$ corresponded to higher A 
values (Figure~9).   This can be explained if the galaxy has had mass accreted 
from the IGM or a merger, disrupting its structural appearance but with 
V$_{\rm max}$ not yet settled.   There are no positive 
deviations in M$_{*}$ larger than $\sim 10^{11}$ \solm, which is 
approximately 80\% of the stellar mass of the Milky Way. This is 
consistent with the idea that minor mergers are the cause of the 
deviations.  Figure~9 also shows that deviations at lower M$_{*}$ 
correspond to lower $A$ values. This could be explained if these 
galaxies have not fully evolve to their final stellar mass.

\section{Summary}

We present in this paper a structural analysis  of 91 disk galaxies 
within the redshift range $0 < z < 1$ to determine if the scatter
in the stellar mass Tully-Fisher (SMTF) relation at $z < 1$ is produced
through mergers or accretion. Previous analysis of the kinematic 
(V$_{\rm max}$) and stellar mass (M$_{*}$) 
properties of these galaxies show no significant evolution in the zero 
point or scatter of the SMTF relation out to $z \sim 1$ 
(Conselice et al. 2005a). There is however a
significant amount of scatter in the SMTF relation
whose origin is unknown.

We investigate whether deviations from the stellar mass Tully-Fisher 
relation, as seen by 
Conselice et al (2005a), can be explained by the presence of 
morphologically disturbed galaxies whose CAS structural parameters are
measured.  We find that on average galaxies which deviate the most from
the stellar mass Tully-Fisher relation are asymmetric systems.  This
correlation is significant at $\sim 3$ $\sigma$, and is not
produced by star formation or edge-on galaxies.  We interpret this to
imply that disk galaxies at $z < 1$ are forming hierarchically either
through accretion of satellite galaxies, or intergalactic baryons and
dark matter.
This is furthermore seen through how deviations occur from the stellar
mass Tully-Fisher relation, as $\sim 70$\% of our disks at $z > 0.7$ 
deviate above the $z \sim 0$ relation.  Growth onto the stellar mass
Tully-Fisher relation can in this case only occur through kinematic
cooling to raise the value of V$_{\rm max}$, or through the addition of
matter, without significant star formation, through accretion or merging.  
We see a significant positive correlation between both 
V$_{\rm max}$ and M$_{*}$ deviations with asymmetry, suggesting that both 
could be  evolving simultaneously. 

Our statistical analysis also suggests that more disk galaxies with $A > 0.25$
are present at high redshift. Both of these results support the hierarchical 
model for disk galaxy formation at $z < 1$, even after the major
epoch of merging has ended and the major properties of disks are in
place (Conselice et al. 2003; Jogee et al. 2004; Conselice et al. 2005b; 
Ravindranath et al. 2006).
Within our galaxy sample, we furthermore find that 70\% of the most
asymmetric galaxies are at $z > 0.4$, implying strong dynamical evolution in 
the last 5 Gyrs.   This is 
further supported when observing the scale of deviations in stellar
mass, which are typically that of a low-mass galaxy.  Our
primary conclusion is that while disk galaxies are morphologically
present in abundance at $z < 1$, these systems are still forming hierarchically
during this time due to collisions with existing galaxies, or discrete
baryonic/dark matter dominated clouds.

There are a number of improvements which could enhance our understanding
of this problem.  This is necessary since both 
CAS parameters, and the method of estimating V$_{\rm max}$ and 
M$_{*}$, have their limitations. Few low stellar mass galaxies are present 
in our sample due to the original selection criteria, and the fact that 
more distant low mass galaxies are harder to measure rotation curves for. 
We therefore do not have a completely representative sample of disk galaxies 
over the redshift range $0<z<1$.  For galaxies within our sample, 
V$_{\rm max}$ may also 
have been miss-estimated due to observational errors in the ground-based slit 
spectroscopy. The method we use is not sufficient to imply detailed interior 
kinematics of a galaxy. However, combined with knowledge of their luminosity, 
it is sufficient to recognise those that do not have well formed disks. In 
conclusion, we find that positive and negative deviations from the $z \sim 0$ 
TF relation seen by Conselice et al (2005a) are indicative of disturbed 
galaxies.   Whilst this is a step towards understanding the origin of the TF
relation, detailed IFU spectra are needed to confirm this, and to determine 
exactly what mechanism causes this disruption. 

We thank Nicole Vogt for useful advice and the University of Nottingham
for their support.

\appendix

\label{lastpage}

\end{document}